\documentclass{PoS}

\def\gsim{\mathrel{\raise.3ex\hbox{$>$\kern-.75em\lower1ex\hbox{$\sim$}}}}

\title{Non-perturbatively improved clover action for SU(2) gauge + fundamental and adjoint representation fermions}

\ShortTitle{Non-perturbative improvement for SU(2)+fermions}

\author{Tuomas Karavirta, Kimmo Tuominen\\
	Department of Physics, P.O.Box 35 (YFL), \\
        FI-40014 University of Jyv\"askyl\"a, Finland, \\
        and \\
        Helsinki Institute of Physics, P.O.~Box 64, \\
        FI-00014 University of Helsinki, Finland\\
	E-mail: \email{tuomas.karavirta@jyu.fi},
        \email{kimmo.i.tuominen@jyu.fi}}

\author{\speaker{Anne Mykkanen}, Jarno Rantaharju$^*$ and Kari Rummukainen\\
	Department of Physics and Helsinki Institute of Physics,\\
	P.O.Box 64, FI-00014 University of Helsinki, Finland\\
	Email: \email{anne-mari.mykkanen@helsinki.fi},
               \email{jarno.rantaharju@helsinki.fi},
               \email{kari.rummukainen@helsinki.fi}}

\abstract{%
  The research of strongly coupled beyond-the-standard-model theories
  has generated significant interest in non-abelian gauge field
  theories with different number of fermions in different
  representations. Motivated by the increased interest to various
  technicolor scenarios, we study the non-perturbative improvement of
  the Wilson-clover action with $SU(2)$ gauge fields and $2$ flavors
  of fermions in the fundamental and adjoint representations.  The
  Sheikholeslami-Wohlert coefficients are fixed using Schr\"odinger
  functional boundary conditions.  The adjoint representation theory
  is a candidate for a "minimal technicolor" theory, already studied
  on the lattice using unimproved Wilson fermions.
}

\FullConference{The XXVIII International Symposium on Lattice Field Theory, Lattice2010\\
		June 14-19, 2010\\
		Villasimius, Italy}

\begin{document}

\section{Introduction}
Gauge theories with fermions in non-fundamental representations have
recently been proposed as candidates for phenomenologically viable
Technicolor models.  A particularly simple example of these is
so-called {\em minimal walking technicolor} (MWT), a theory with
SU(2) gauge fields and two adjoint representation fermions
\cite{Sannino:2004qp,Dietrich:2005jn,Dietrich:2006cm}.  Analytical
studies suggest that this theory has either an infrared stable fixed
point or the coupling constant ``walks,'' i.e. evolves very slowly
over some energy range.  This question is inherently non-perturbative
and lattice simulations are needed.

Initial lattice studies of MWT suggest that the theory has an infrared
stable fixed point
\cite{DelDebbio:2010hx,DelDebbio:2010hu, Hietanen:2008mr, Hietanen:2009az, Catterall:2007yx,Catterall:2008qk,DelDebbio:2008zf}.
However, these investigations were made with non-improved Wilson
fermion action, which is known to have large $O(a)$ cutoff effects.
The cutoff effects can be especially significant when the evolution of
the coupling constant is measured with the Schr\"odinger functional
method \cite{DelDebbio:2010hx, Hietanen:2009az}; where, with the
unimproved action, it was not possible to perform the continuum limit
in a controlled fashion.  Thus, it is important to use actions which
have as small cutoff effects as possible.  This becomes especially
important at relatively large bare lattice couplings, which have been
observed to be necessary in order to study the relevant physical
domain.

Our aim is to calculate the $O(a)$ improvement coefficients
for the Wilson-clover action for MWT, and, for comparison,
also for SU(2) gauge theory with two fundamental fermion
flavours.  We do this in two stages: in this work
we describe the non-perturbative evoluation of the
clover (Sheikholeslami-Wohlert) coefficient in these
two theories using
the Schr\"odinger functional method 
\cite{Luscher:1991wu,Luscher:1992zx,Luscher:1996ug,
  Jansen:1998mx}.  In order to apply the Schr\"odinger
functional for the calculation of the coupling constant,
we also need to calculate various ``boundary improvement''
coefficients.  This we do using perturbation theory, and
the calculation is described in these proceedings in ref.~\cite{sfboundary}.

\section{The Models}

We study the nonperturbative order $a$ improvement of two lattice
gauge models, one where two flavors of fermions couple to the
fundamental representation of an SU(2) gauge field, and one where they
couple to the adjoint representation. They share the same standard
Wilson gauge action for SU(2). The fermion action is
\begin{eqnarray} \label{fermaction}
S_F &=& a^4 \sum_x \bar\psi(x)\left ( D+m_0 \right ) \psi(x) \\
D &=& \frac 12 \left [ \gamma_\mu({\nabla}^*_\mu + 
  {\nabla}_\mu) - a{\nabla}^*_\mu{\nabla}_\mu \right ] + 
c_{\rm SW} \frac{ia}{4}\sigma_{\mu\nu}\hat F_{\mu\nu},
\end{eqnarray}
where $\hat F_{\mu\nu}$ is the symmetrized field strength tensor
(clover) and 
\begin{equation}
  \nabla_\mu \psi(x) = 
  \frac 1 a \left[\widetilde{U}_{\mu}(x)\psi(x+\hat\mu) - \psi(x)\right],
~~~~~~~
  \nabla^*_\mu \psi(x) = 
  \frac 1 a \left[\psi(x) -         
   \widetilde{U}^\dagger_{\mu}(x-\mu)\psi(x-\hat\mu)\right].
\end{equation}
Here $\widetilde U$ is the gauge link in the appropriate fermion
representation, that is, it is the standard fundamental SU(2) link for
fundamental representation fermions, and for adjoint fermions it is
\begin{equation}
\widetilde{U}^{ab}_\mu(x) = 
2 \mbox{Tr\,} \left ( \lambda^a U_{\mu}(x) \lambda^b U_{\mu}^\dagger(x)  \right ), ~~~~~
\lambda^a = \frac 12 \sigma^a, \,\,\,\, a=1,2,3.
\label{Vlinks}
\end{equation}

\section{Schr\"odinger functional scheme}

Our method for determining the improvement coefficient $c_{\rm SW}$
follows refs.~\cite{Luscher:1992zx,Luscher:1996ug,Jansen:1998mx}, 
where the Schr\"odinger
functional scheme is used to determine $c_{\rm SW}$ in the case of $QCD$.
However, for the adjoint representation fermions (and, in general,
for fermions in higher representations) the method
is modified, as described below.

We shall work with lattices of size $L^3 \times T$.  In the
Schr\"odinger functional scheme the spatial gauge fields are fixed to
constant values at time slices $x_0=0$ and $x_0=T$, chosen so that these
generate a chromoelectric background field. 

For fundamental fermions we use color diagonal background fields
as in ref.~\cite{Luscher:1992zx}
\begin{eqnarray} 
  U_k(x_0=T) &=& \exp(i C'), \,\,\,\, 
  C' = -\frac{\pi}{4} \frac{a\sigma^3}{L} 
  \label{csw_abelian_boundary1} \\
  U_k(x_0=0) &=& \exp(i C), \,\,\,\,
  C = -\frac{3\pi}{4} \frac{a\sigma^3}{L}.
  \label{csw_abelian_boundary2}
\end{eqnarray}
These generate a chromoelectric background field $\propto \sigma^3$.
Different boundary conditions give rise to different
cutoff effects in fermion propagation when the source is
at $x_0 = 0$ or at $x_0=T$.  The idea is to find the
value of $c_{\rm SW}$ which maximizes the symmetry between
the two cases, leading to automatic $O(a)$ improvement.



For adjoint representation fermions, however, complications
emerge. Using Eq.~(\ref{Vlinks}) we immediately 
notice that the boundary matrices 
(\ref{csw_abelian_boundary1}), (\ref{csw_abelian_boundary2})
are transformed to form
\begin{displaymath}
  \widetilde{U}_{k} =
  \left( \begin{array}{ccc}
      \ldots & \ldots & 0 \\
      \ldots & \ldots & 0 \\
      0 & 0 & 1
    \end{array} \right)
\end{displaymath}
%
Thus, there is a component of the adjoint fermion spinor which simply
does not see the background field.  This feature is independent
of the color structure chosen for the boundary conditions.
It turns out that regardless
of how the fermion sources or the constant boundary conditions are chosen, 
at long distances the fermions propagate as if there is no
background field.    In other words,
the adjoint fermions ``see'' the background electric field only at 
short distances.  

This property gives the 
background field method significantly less leverage for determining
$c_{\rm SW}$ for adjoint representation fermions.  In order to 
maximize the effect of the different boundaries, we 
choose to maximize the difference between the two boundaries,
and we use the following asymmetric "non-Abelian" boundary 
conditions: links at the upper 
$x_0=T$ boundary are chosen to be trivial
\begin{equation} \label{csw_upper_boundary}
U(x_0=T,k) = I
\end{equation}
and at the lower boundary $x_0=0$ we use
\begin{equation} \label{csw_lower_boundary}
U(x_0=0,k) = \exp(a C_k), \,\,\,\, C_k = \frac{\pi}{2} \frac{\tau^k}{i L}.
\end{equation}
These boundary conditions do not make the problem to vanish, but ameliorate it 
to a degree.  We also note that these boundary conditions are 
useful only for determining $c_{\rm SW}$, not for evaluating the
coupling constant.  


We define the fermion mass through the partial conservation of the axial current (PCAC) relation:
\begin{equation}
  m_Q(x_0)=\frac{1}{2}\frac{\frac{1}{2}(\partial_0^\ast+\partial_0)f_A(x_0)+c_A a\partial_0^\ast\partial_0 f_P(x_0)}{f_P(x_0)}\equiv r(x_0)+c_As(x_0),
\label{pcac}
\end{equation}
where
\begin{eqnarray}
A^a_\mu &=& \bar{\psi}(x)\gamma_5\gamma_\mu\frac{1}{2}\sigma^a\psi(x), \\
P^a &=& \bar{\psi}(x)\gamma_5\frac{1}{2}\sigma^a\psi(x),\\
f_A(x_0) &=& -a^6\sum_{\bf{y,z}}\langle A_0^a(x)\bar{\zeta}({\bf{y}})\gamma_5\frac{1}{2}\sigma^a\zeta({\bf{z}})\rangle,
\label{fa}\\
f_P(x_0) &=& -a^6\sum_{\bf{y,z}}\langle P^a(x)\bar{\zeta}({\bf{y}})\gamma_5\frac{1}{2}\sigma^a\zeta({\bf{z}})\rangle.
\label{fp}
\end{eqnarray}
Here the sources $\zeta(\bf{z})$ live on the time slice $x_0 = 0$.
The term proportional to $c_A$ in Eq.~(\ref{pcac}) is irrelevant in the continuum limit, but it is needed to cancel $O(a)$-contributions to the axial current in the Wilson action.  

Analogously with Eqs.~(\ref{fa},\ref{fp}), we define
another set of correlation functions, $f_A^\prime(T-x)$, $f_P^\prime(T-x)$ and $r'(T-x)$, $s'(T-x)$, where the source is now at time slice $x_0=T$.  
In order to obtain an expression which is independent of $c_A$
we consider the combination \cite{Luscher:1996ug}
\begin{equation}
  M(x_0,y_0)=r(x_0)-s(x_0)\frac{r(y_0)-r^\prime(y_0)}{s(y_0)-s^\prime(y_0)},
  \label{M}
\end{equation}
which coincides with $m_Q$ up to ${\mathcal{O}}(a^2)$ corrections.
In our calculations here we define the fermion mass with 
$m = M(T/2,T/4)$.

Further defining $M^\prime$ with obvious replacements to (\ref{M})
gives us two correlation functions which, in the absence of cutoff
effects, are equal.  Thus, the quantity
\begin{equation}
  \Delta M\equiv
  M(\frac{3}{4}T,\frac{1}{4}T) - 
  M^\prime(\frac{3}{4}T,\frac{1}{4}T)
  \label{deltaM}
\end{equation}
vanishes up to corrections of ${\mathcal{O}}(a^2)$ if $c_{\rm SW}$ has its proper value.\footnote{%
  To be more precise, at tree level ($g=0$, $c_{\rm SW}=1$) 
  both $\Delta M$ and $m=M(T/2,T/4)$ have small
  values, which depends on the boundary
  conditions and the lattice size.  In order to obtain the correct
  weak coupling limit We actually match $\Delta M$
  and $m$
  to these tree-level values, not to zero 
  (see ref.~\cite{Luscher:1996ug}).}

In order to achieve full $O(a)$ improvement in the Schr\"odinger
functional schema we need to cancel also the $O(a)$ errors caused by
the fixed boundaries.  We treat these by using 1-loop improved
boundary conditions, discussed in ref.~\cite{sfboundary}.  These are
not necessary, however, for the calculation here.

\section{The simulations and results}

In order to evaluate $c_{\rm SW}$ we used the following routine: we
choose lattice volume $L^3\times T = 8^3\times 16$ for both
fundamental and adjoint representation fermions, and a set of values
of the lattice coupling $\beta$.
\begin{enumerate}
\item For a given $\beta$, we choose initial $c_{\rm SW}$ (typically
  extrapolating from results obtained with previous values of
  $\beta$).

\item We choose a couple of values for $\kappa=\frac{1}{8+2am_{0}}$,
  and determine by interpolation the critical value
  $\kappa_c(\beta,c_{\rm SW})$ where the fermion mass $M(T/2,T/4)$
  vanishes.
  
\item Once we have an estimate of the critical $\kappa$, we choose a
  new value for $c_{\rm SW}$ and repeat the search of $\kappa_c$.

\item At the same time, we measure $\Delta M(c_{\rm SW})$.  Now we can
  linearly interpolate/extrapolate in $c_{\rm SW}$ so that $\Delta M$
  vanishes, obtaining the desired value of $c_{\rm SW}(\beta)$.
  Using simulations at this final $c_{\rm SW}$ we can relocate the critical
  $\kappa$, if desired, and verify the results of the interpolation.
\end{enumerate}

\begin{figure}
  \centerline{\includegraphics[width=.6\linewidth]{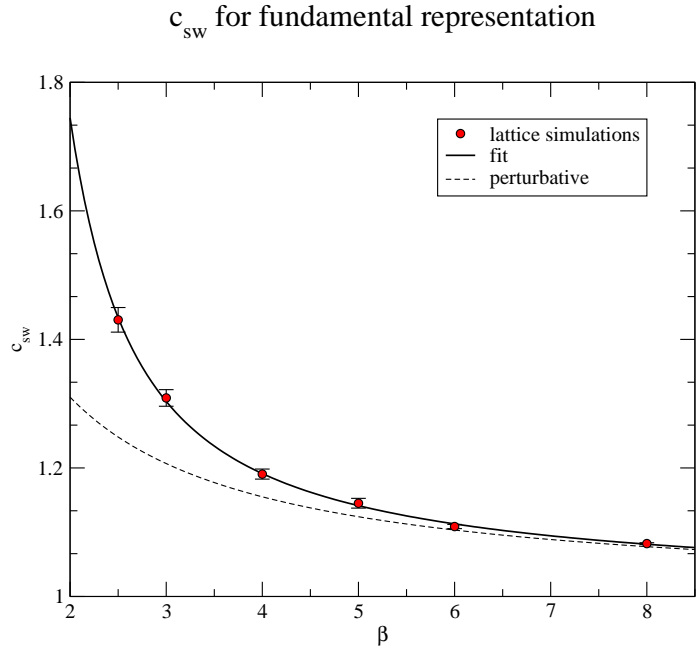}}
  \caption[a]{$c_{\rm SW}$ for two flavors of fundamental 
    representation fermions.  The solid line is the
    interpolating fit, Eq.~(\ref{fundfit}), and
    the dashed line is the 1-loop perturbative
    value}
  \label{fig:fund}
\end{figure}

\begin{figure}
  \centerline{\includegraphics[width=.6\linewidth]{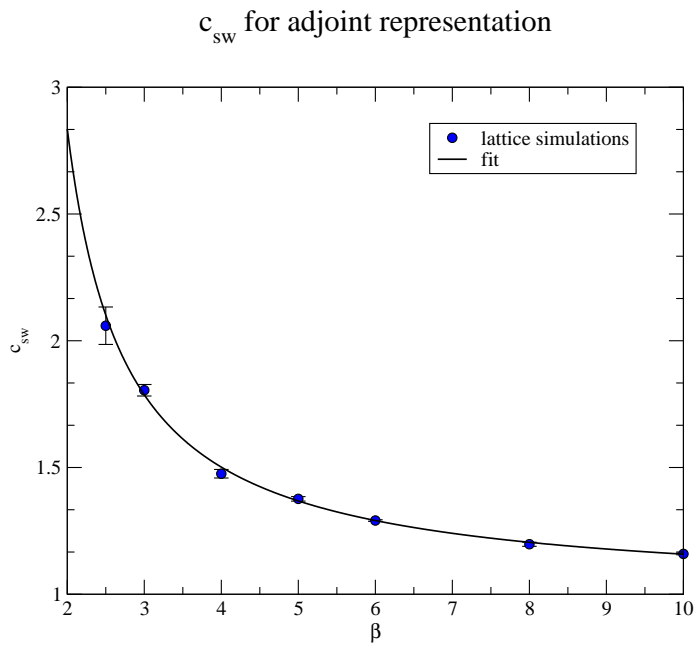}}
  \caption[b]{$c_{\rm SW}$ for two flavors of adjoint 
    representation fermions, with
    the interpolating fit, Eq.~(\ref{adjfit}).}
  \label{fig:adj}
\end{figure}

In Figs.\ref{fig:fund} and \ref{fig:adj} we show our results for the 
clover coefficient $c_{\rm
  SW}$ for both fundamental and adjoint representations. 
The values of $\beta$ used are 
$\beta = 2.5, 3, 4, 5, 6, 8$, and also $\beta=10$ for
the adjoint representation.

Finally, the measured values for $c_{\rm SW}$ can be fitted 
with a rational interpolating expression, which can used 
in simulations for this range of $\beta$-values.  
For fundamental representation fermions we use the
perturbative 1-loop result 
  $c_{\rm SW} = 1 - 0.1551(1) g^2 + O(g^4)$
\cite{Luscher:1996vw} to constrain the fit:
\begin{equation}
c_{\rm SW,fund}= 
\frac{1-0.090254g^2-0.038846g^4+0.028054g^6}{1-(0.1551+0.090254)g^2}.
\label{fundfit}
\end{equation}
For the adjoint representation the perturbative result is not known,
and we obtain the fit result
\begin{equation}
c_{\rm SW,adj} = 
\frac {1+0.032653 g^2 -0.002844 g^4}{1-0.314153 g^2}.
\label{adjfit}
\end{equation}
In both cases the interpolating fits are valid for $\beta \gsim 2.5$.
For the adjoint fermions it is difficult to reach smaller $\beta$-values
because $c_{\rm SW}$ grows rapidly; and while we were able to reach 
$\beta=2.3$ the errors were too large to constrain the fit (\ref{adjfit})
further.

\acknowledgments
This work is supported by the Academy of Finland grant 1134018.
J.R is supported by V\"ais\"al\"a Foundation, A.M. is supported by Magnus Ehrnrooth Foundation.
The simulations were carried out at
the Finnish IT Center for Science (CSC) and at the Institute for
Development and Resources in Intensive Scientific computing (IDRIS).

\end{document}